IAC-22-A3.3B

# Feasibility assessment of optical communications between ground and satellite on Mars through the simulation of atmospheric effects on signal quality leading to a proposal for a new communications network architecture during extreme weather

**Zachary Rowland[a], Eva Fernández-Rodríguez[b]**

[a] *Aerospace Engineer*
[b] *Telecommunications Engineer*

**Abstract**

Mars is the next milestone in human exploration. However, there are still several challenges that must be assessed to ensure appropriate conditions in a future settlement. Communications services will be essential for this task, providing not only a link between Earth and Mars but also supporting Martian weather forecasting and any potential rescue missions. These applications require a robust, high data rate communications network that allows for rapid response, remote sensing and public engagement.

This research aims to study the feasibility of ground-to-satellite (and vice versa) optical communications during extreme Martian weather conditions, focusing on the link between a ground station on the surface of Mars and a satellite orbiting the planet. Long-lasting and expansive Martian dust storms, particularly common in the southern hemisphere, pose a considerable challenge when considering the feasibility of optical communications with Mars due to their significant impact in terms of signal attenuation and scattering.

The methodology of this study is based on a computer simulation of the system featuring the characterisation of the Martian atmosphere and optical link to measure the attenuation and undesired effects suffered by the data signal when applying different environmental configuration parameters. The flexibility of the approach allows for the prediction of communications link quality in extreme cases such as global dust storms. The simulation is based on atmospheric data from the Mars Reconnaissance Orbiter's Mars Climate Sounder instrument and considers the recently launched Laser Communication Relay Demonstration (LCRD).

The extreme conditions during dust storms in the southern polar-hood region lead to the proposal of a new communications network architecture to ensure connectivity during these events. The proposed operation involves the detection of heavy signal attenuation and triggers a two step communications link assisted by a UAV for data relay.

The outcomes of the study may be used by future missions to evaluate the feasibility of optical communications with the planet's surface in different Martian weather conditions.

**Nomenclature**

| | |
|---|---|
| $P_r$ | Power at the receiver |
| $P_t$ | Power at the transmitter |
| $A$ | Receiver aperture area |
| $\theta$ | Divergence angle |
| $L$ | Distance travelled by the beam |
| $\gamma$ | Total atmospheric attenuation factor |
| $\tau_t$ | Transmitted efficiency |
| $\tau_r$ | Received efficiency |
| $S$ | Sensitivity |
| $Q_{ext}$ | Extinction efficiency factor |
| $Q_{sca}$ | Scattering efficiency factor |
| $Q_{abs}$ | Absorption efficiency factor |
| $\sigma_{sca}$ | Particle cross section |
| $R$ | Particle radius |
| $\rho$ | Particle concentration |

**Acronyms/Abbreviations**

| | |
|---|---|
| ATP | Acquisition, Tracking and Pointing |
| FSO | Free Space Optical |
| MCS | Mars Climate Sounder |
| MRO | Mars Reconnaissance Orbiter |
| MY | Mars Year |
| UAV | Unmanned Aerial Vehicle |
| VTOL | Vertical Take-Off and Landing |
| LCRD | Laser Communications Relay Demonstration |
| EDRS | European Data Relay System |

## 1. Introduction

To date, there have been 10 successful landings of probes or rovers on the surface of Mars (not including sub-landers such as the Mars helicopter Ingenuity) [1], with all but one (Zhurong, landed in 2021) launched and operated by NASA. Radio communication with NASA's landers is either direct with Earth, or is relayed via orbiters in Martian orbit. For example, NASA JPL's

Curiosity rover achieves the following data rates using these two methods:

Table 1. Data rates for Curiosity Rover [2]

| Method | Minimum data rate (bps) | Maximum data rate (bps) |
|---|---|---|
| Direct-to-Earth | 500 | 32,000 |
| MRO* | - | 2,000,000 |
| Mars Odyssey | 128,000 | 256,000 |

* MRO = Mars Reconnaissance Orbiter

An orbiter is in position to communicate with the rover for around eight minutes per sol, allowing between 100 - 250 megabits of data to be transmitted to an orbiter during that period. The data rate for MRO to NASA's Deep Space Network antennas on Earth is around 0.5 - 4 mbps [3].

In comparison, free-space optical (FSO) communications can achieve data rates on the order of hundreds of Gbps [4]. Therefore, as Mars exploration continues into the future there is a clear advantage in terms of data rate for the utilisation of optical over radio communication with Earth. However, much of the existing research on optical communications in space explores only either transmission between spacecraft in deep space or between spacecraft and ground stations on Earth. Such a system would still be limited in terms of data rate between the Martian surface and the relay spacecraft in orbit around Mars, if this step is still carried out via radio link. Given the limited time window for transmitting to an orbiter per sol, this presents a significant bottleneck.

Therefore, this paper explores the feasibility of optical communications between the surface of Mars and a Martian orbiter. The higher data rates made available through such a system could be essential for establishing a human presence on Mars due to the need to transmit significantly larger amounts of data for scientific, personal and emergency communications. This is done primarily through literature review and by analysing atmospheric and dust storm data collected by the MRO Mars Climate Sounder (MCS) instrument to provide input to calculations of attenuation from optical communications in a surface-to-satellite laser link. This model will ultimately be used to develop a code block for GNU Radio to aid future designs for optical systems on Mars [5].

Finally, we explore a new communications network architecture to ensure connectivity during these events, based on a two step communications link assisted by a UAV for data relay.

## 2. Literature review
### 2.1. The Martian atmosphere

Mars has a thin (relative to Earth) atmosphere consisting of (as measured by Curiosity at the surface of Gale crater) 95% by volume of carbon dioxide ($CO_2$), 2.6% molecular nitrogen ($N_2$), 1.9% argon (Ar), 0.16% molecular oxygen ($O_2$), and 0.06% carbon monoxide (CO). However, there are variations in this composition throughout the Martian seasons. [6]

Mars also differs from Earth in a number of other key respects that impact its atmosphere, notably: the lack of a biosphere, the lack of a permanent presence of liquid water on the surface, the smaller size of Mars, the greater distance of Mars from the sun, and the lack of a core-generated Magnetic field (Mars instead has a magnetosphere induced by solar winds that may have stripped away a much thicker atmosphere in the past [7]). Mars also lacks any real stratosphere due to the low concentration of ozone in the atmosphere (driven by the low concentration of $O_2$).

Despite these clear differences, there are also similarities between the processes that drive the behaviour of the atmospheres on Mars and on Earth as highlighted in [8]. Temperature variations are driven by radiative heating from the Sun and cooling from thermal radiation to space and these variations are moderated by global atmospheric circulation systems. Both planets have similar rotation rates and obliquity, leading to similar global circulation patterns dominated by direct overturning motions at low latitudes and cyclone-anticyclone weather systems at higher latitudes, with similar seasonal variations between hemispheres.

As on Earth, pressure, temperature and density vary with altitude on Mars. Following a simple model from NASA [9] developed from measurements of the Martian atmosphere made by the Mars Global Surveyor for educational purposes gives the following variation in altitude of each of these parameters (Fig. 1). However, when compared with measurements from MRO MCS (Fig. 4 and Fig. 6), it can be seen that temperature actually fluctuates with altitude and is quite dependent on weather conditions, similar to on Earth (as can be seen in the International Standard Atmospheres) [10].

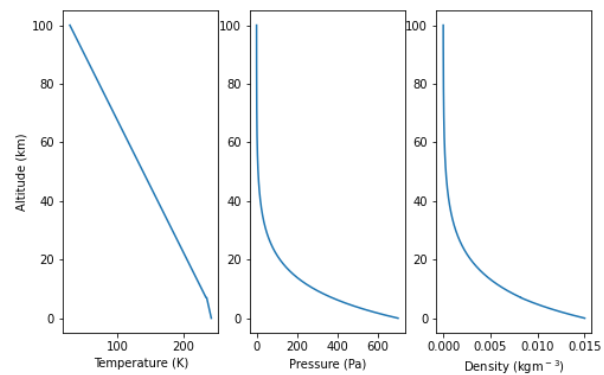

Fig. 1. Variation in Temperature, Pressure and Density with Altitude according to simple NASA Mars Atmosphere Model

In clear conditions (Fig. 6), the concentration of dust particles in the atmosphere decreases predictably with altitude, however during dust storms particulates can be lifted from the ground increasing the concentration of dust in the upper atmosphere (Fig. 4). This can be driven by seasonal variations in the amount of radiative energy received from the Sun, which increases the temperature and wind speeds that lift dust from the surface and into the atmosphere [11].

The quality of optical communications on Mars would be largely driven by the same factors that drive radiative forcing on the planet, particularly where it concerns the visible and infrared parts of the spectrum. [8] The high in $CO_2$ gas mixture on Mars is relatively transparent to visible light and only slightly absorbing in the near-infrared. However, $CO_2$ absorbs strongly in the thermal infrared at Martian temperatures (due to absorption bands centred around 15, 4.3 and 2.7 μm), causing around 20% of thermal radiation emitted from the surface to be absorbed. This difference in opacity leads to a significant greenhouse effect on Mars. This could similarly affect optical communications, as $CO_2$ also has absorption bands in the near-infrared (1.4, 1.6, 2.0, 2.7, 4.8, 5.2 μm) [12], with the lowest wavelength bands being of relevance to the choice of signal wavelength. However the significance of this effect is dependent on atmospheric composition, as the absorption coefficient varies with pressure and temperature [13] (see also section 3).

Aside from the atmospheric absorption, there is some effect from absorption and scattering due to suspended particulares in the Martian atmosphere, either dust particles lifted up from the surface or condensate cloud particles ($H_2O$ or $CO_2$ ice). The radiative effects of these particles depend on various factors discussed further in section 3.

Water and ice clouds are typically optically thin [8], however they can affect FSO communications with Mars in two ways: $H_2O$ ice clouds are highly reflective in the visible part of the spectrum with an albedo close to 1, leading to scattering in the visible; and $CO_2$ ice clouds may scatter across the entire visible and infrared spectrum due to particle sizes ranging from 10-100 μm in well developed clouds. The optical depth of water ice clouds is often anticorrelated with that of dust storms, as dust storms form largely during the perihelion season (Ls = 180° - 360°), while the greatest extent of water ice clouds are observed during the colder aphelion season (Ls = 0° - 180°) [14]. The effect of clouds on absorption and scattering of signals is not addressed in this study at present, but is of interest for future work (see section 5).

*2.2. Designs for UAVs on Mars*

One possibility for mitigating losses during dust storms on Mars is to transfer data to an Unmanned Aerial Vehicle (UAV), which can operate above the thickest part of the atmosphere and continue to transmit data during the event. Such an approach could be particularly useful when the data being transmitted has been collected by the UAV itself, such as during atmospheric remote sensing missions, as this would reduce dependence on a home base of operations for data transmission. The potential for UAV use on Mars has been demonstrated recently with the Mars Helicopter (Ingenuity) technology demonstration, with a current record flight time of 166.4 seconds [15]. A rotorcraft design was also selected for NASA's New Frontiers Program with the Dragonfly mission to Titan, where a dual-quadcopter UAV will be capable of travelling 10 m/s at maximum-range speed (according to flight performance analysis) [16].

A significant challenge for high-altitude flight on Mars is the low atmospheric pressure, even at low altitudes. This makes generating lift much more difficult than on Earth (atmospheric pressure at the Martian surface is only 0.7% of Earth at sea level), necessitating much larger or faster rotating lifting surfaces (in the case of fixed-wing, glider or rotorcraft UAVs) or relatively greater hull volumes or internal gas temperatures in the case of a balloon craft.

In [17], the authors consider several different use cases for UAVs on Mars, as well as the challenges associated with operating a UAV in the Martian atmosphere. Flight of a prototype UAV is simulated on Mars. The designs considered in their preliminary studies include airships and balloons, VTOL (vertical take-off and landing) aircraft, gliders, and flapping wing aircraft.

Balloons are particularly interesting for the data-relay use case as they can fly unpowered at higher altitudes than other forms of aircraft, while maintaining a relatively static position if tethered or otherwise controlled (at least when compared with fixed-wing or glider craft).

The concept of a Martian balloon UAV has been investigated for many years. For example, Mars 2001 Aerobot was a study initiated at NASA Jet Propulsion Laboratory in 1995 to target the 2001 Mars launch opportunity, deploying a balloon with a target cruise altitude of 5 - 8 km above reference level [18].

The concept of a Mars boundary layer sounding balloon was more recently investigated in [19], specifically for a balloon launched off the payload deck of a lander from the Martian surface. This concept was simulated with a launch from a Jezero crater landing site with a flight time of 50 hours, reaching a height above ground of 4 km (corresponding to 2 km above the reference geoid). Some physical characteristics of the design simulated are included in Table 2 below (not including the mass of the proposed support system assembly or pressure vessel). The total mass of the balloon flight element is approximately 2.3 kg.

Table 2. Some physical characteristics of the balloon-system proposed in [19]

| Characteristic | Value |
| --- | --- |
| Hull diameter | 7 m |
| Hull avg. thickness | 5.5 μm |
| Hull mass | 1450 g |
| Gondola mass | 600 g |
| Mass of lifting gas | 300 g |

Achieving an altitude of 2 km above the reference geoid would allow the UAV to operate at an atmospheric density roughly 15% lower than that of the surface (from 0.015 to 0.013 $kgm^{-3}$, following the model in Fig. 1). By increasing the balloon diameter to 7.5 m (and keeping the gondola mass constant at 600 g), the balloon could operate at roughly 3.3 km, leading to a 25% reduction in density (to 0.011 $kgm^{-3}$). Stratospheric balloons on Earth operate at altitudes up to 42 km [20], corresponding to an atmospheric density of as little as 0.003402 $kgm^{-3}$ (interpolated from the US Standard Atmosphere in [21]). The same density would be expected at an altitude of approximately 18 km on Mars, although ballooning altitude would still be further limited by the cold atmosphere of Mars, particularly at night [22]. Nonetheless, aside from the clear limitation of packing constraints for a mission, there is potential to design a balloon capable of operating well above the densest part of the Martian atmosphere in the right conditions.

There are however several apparent problems with adapting such a balloon to carry an optical transceiver for data relay, specifically:

1. The mass of the transceiver;
2. The reality that most practical balloon designs consider a gondola carried below the hull, which would block the transciever's line of sight when the satellite is directly overhead;
3. The relative difficulty in maintaining a lock on the satellite position with a transceiver onboard a moving gondola vs. on the ground;
4. As shown in [23], during Martian dust storms dust particles may be lifted so high that the greatest particle concentrations are still above any practical altitude for balloon operation; and
5. Balloons are essentially uncontrollable without onboard thrust control (this may be restricted to only altitude control if taking adequate advantage of wind patterns at different altitudes) or a long tether attaching them to the launch site, both of which could add significant mass and complexity to the design. On Earth, the record for tethered balloon flight is 7 km altitude [24].

Because of these issues and the relatively low opacity of the Martian atmosphere (when compared with Earth, see section 4.2), we conclude that such a system may be unlikely to be considered practical or necessary in the near future. If these issues were to be addressed, a system could however be envisioned where, on detection of heavy signal attenuation due to an atmospheric event:

1. Data is uploaded to storage on the gondola via a physical connection at a base station;
2. The balloon is launched in conditions of poor visibility, while being connected to the station with a tether;
3. Upon reaching maximum altitude, the transceiver onboard the gondola would maintain a lock on the satellite and transmit data during each pass (or continuously in the case of a satellite in areostationary orbit);
4. The balloon would then be reeled back to ground as needed to recharge systems and replace lost gas to the hull.

This system is summarised in Fig. 2 below.

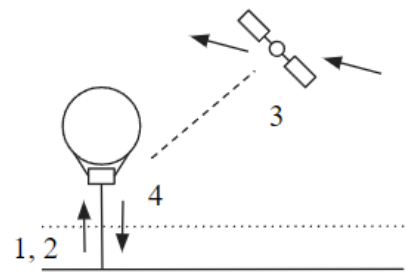

Fig. 2. A simple design for a two step data communications link assisted by a UAV for data relay (the dotted line above the ground represents the boundary of highest atmospheric and/or dust density)

*2.3. The Laser Communications Relay Demonstration*

The Laser Communications Relay Demonstration (LCRD) is a NASA technology demonstration mission launched in December 2021, with the aim of acting as a relay for future missions to send data to specific ground stations on Earth using optical communications. The ground stations for this system are located strategically on Earth to avoid loss of visibility due to cloud cover which would block light transmission and, consequently, decrease the performance of the optical link. The LCRD capabilities make it the first NASA two-way, end-to-end optical relay [25].

Other demonstrations such as the European Data Relay System (EDRS) launched in January 2016 also showcase optical communication relays. The purpose of the EDRS is to transmit large quantities of data from other missions down to Earth ground stations in near real time. The European system allows for optical and Ka-band communications links between satellites, aircraft/UAVs or Earth ground stations [26, 27].

From May 2022 NASA's LCRD is prepared to conduct experiments testing and refining the long-term effect of the atmosphere on laser communications signals, assessing the technology applicability and testing the relay capabilities [28]. While the conclusions from these tests are not mature enough at the moment, the results of the LCRD mission will be relevant for this study due to the practical insights for optical communications they will provide in the near future.

## 3. Martian optical communications model

The main advantages of FSO or Laser communications over other forms of wireless communication are higher bandwidth and consequent high data rates, reduced power and mass requirements, high directivity, security and robustness to electromagnetic interference, and the lack of licensing requirements. Despite these upsides, FSO development is challenging due to sensitivity to weather conditions and atmospheric turbulence-induced fading, as well as tight requirements for the acquisition, tracking and pointing (ATP) system due to the narrow beam divergence.

The optical communications model must consider the main factors that affect the optical transmitted power before reaching the receiver. These effects are system loss, geometric loss, misalignment loss, atmospheric loss, atmospheric turbulence induced fading, and background noise. For the case of Mars, particular atmospheric specification is necessary due to the differences in weather and climate in comparison with the Earth.

Particles and gas molecules in the atmosphere cause optical power attenuation due to absorption and scattering of light. The scattering coefficient becomes high for near infrared optical communications when the particle diameter is on the order of the wavelength and may seriously affect the signal quality (see section 2.1). The approach in this study follows the characterization included in [29] and it considers a total light attenuation represented by gaseous absorption and martian dust aerosol scattering coefficients.

Atmospheric turbulence is caused by inhomogeneities in the temperature and pressure along the propagation path. This leads to the generation of turbulent cells, also called eddies, of different sizes and refractive indexes [30]. The variations in the refractive index of the atmosphere affect the light propagation and, therefore, the transmission of the optical signal. In the case of a turbulent cell size on the order of the beam size, an effect called scintillation is generated that is characterised by intensity and phase fluctuations of the received signal that degrades the FSO system performance.

The received signal power can be estimated after having considered the effects on the light propagation. The mathematical relation between the received and transmitted power from a laser follows Beer's Law and involves the physical characteristics of the laser as well as the total atmospheric attenuation factor [31]:

$$P_r = P_t \frac{A}{\theta^2 L^2} e^{-\gamma L} \tau_t \tau_r \qquad (1)$$

Where $P_r$ and $P_t$ are the power at the receiver and transmitter respectively, $A$ is the receiver aperture area, $\theta$ is the divergence angle, $L$ is the distance travelled by the beam, $\gamma$ is the total atmospheric attenuation factor in dBkm$^{-1}$, $\tau_t$ is the transmitted efficiency and $\tau_r$ is the received efficiency.

## 4. Analysis and discussion
*4.1. Analysis of Mars Climate Sounder atmospheric data*

Data from the Mars Reconnaissance Orbiter's Mars Climate Sounder is used in this study to provide input conditions to the optical communications calculations. Specifically, the dataset used is from the MRO MCS Derived Data Record (DDR) [32] which consists of profiles of successfully retrieved geophysical parameters, estimated retrieval of observation errors, and the geometry information to properly locate the profiles in space and time. This derived data is generated from the Reduced Data Record (RDR) calibrated data (which itself is generated from the raw Experiment Data Record), and can be downloaded from NASA's Mars Orbital Data Explorer portal [33].

MCS observes the atmospheric limb of the Martian horizon with 21 detectors in each of 9 spectral bands; 8 thermal infrared channels characterise atmospheric temperature, pressure, water vapour and condensates, while the remaining spectral channel primarily characterises the effects of solar radiation on the Martian energy budget. Measurements are centred approximately 5 km apart at the atmospheric limb, recording vertical profiles of temperature, pressure, water vapour, dust and condensates. Combining these profiles allows for the creation of two-dimensional plots of properties with altitude, as well as three-dimensional "maps" of the Martian atmosphere.

For this study, a specific site and season was sampled to compare the simulated effects on the optical signal. Argyre Planitia was selected as the site for two reasons: it is located in the southern hemisphere where dust storms are more prevalent; and being a plain the geography is fairly homogeneous, allowing data to (in theory) be averaged over a large region without obscuring the effects of the topography as significantly as in a mountainous region. Furthermore, Argyre Planitia has yet to be visited by any Mars surface landers despite the scientific value of its theorised past habitability, making it a contender for future missions and therefore a site of interest for this study [34].

Data from this site (altitude of the pressure surface above the surface point, $h$; temperature, $T$; pressure, $P$) was collected in two sample time periods, to consider conditions during a dust storm and in clear conditions:

1. 01 - 05 July 2018 (MY 34 $L_s$ 203° - 205°), during the peak of the 2018 Mars global dust storm [35].
2. 01 - 05 October 2019 (MY 35 $L_s$ 87° - 89°), during a period of lower atmospheric opacity as measured by NASA InSight in Elysium Planitia [36].

The mean average of profile temperature and pressure was then averaged over four different altitude ranges to acquire representative values for those parts of the atmosphere: the full range of data available, 0 - 40 km, 0 - 20 km and 0 - 15 km (Table 3 and Table 4). Sample profiles are also included from both scenarios as examples (Fig. 4 and Fig. 6).

For the rest of the study the mean values from all profiles from the 0 - 20 km ranges were used for each scenario, as this region covers the bulk of the atmospheric mass.

Table 3. MRO MCS data averaged over different altitude ranges during MY 34 $L_s$ 203° - 205° (dust storm)

| Profile | $h$ [km] | $\overline{T}$ [K] | $\overline{P}$ [Pa] |
|---|---|---|---|
| All profiles | Full range | 182.325 | 49.168 |
| | 0 - 40 | 205.378 | 155.86 |
| | **0 - 20** | **197.399** | **269.26** |
| | 0 - 15 | 195.842 | 317.55 |
| Sample profile * | Full range | 184.363 | 54.076 |
| | 0 - 40 | 187.289 | 159.26 |
| | 0 - 20 | 193.758 | 280.55 |
| | 0 - 15 | - | 336.08 |

* **Id 266:** lat, lon: -45.688, -48.909; $L_s$: 205.29945°

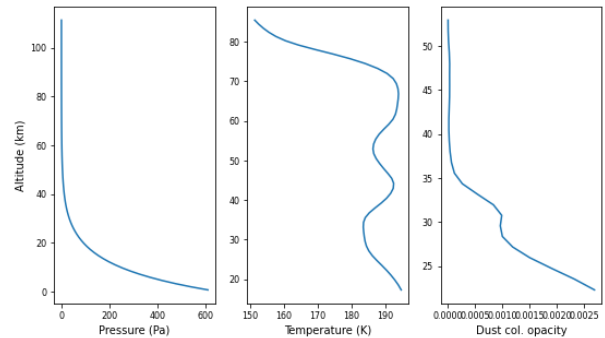

Fig. 4. Pressure, temperature and dust column opacity per altitude of the pressure surface above the surface point for sample profile Id 266 (during dust storm)

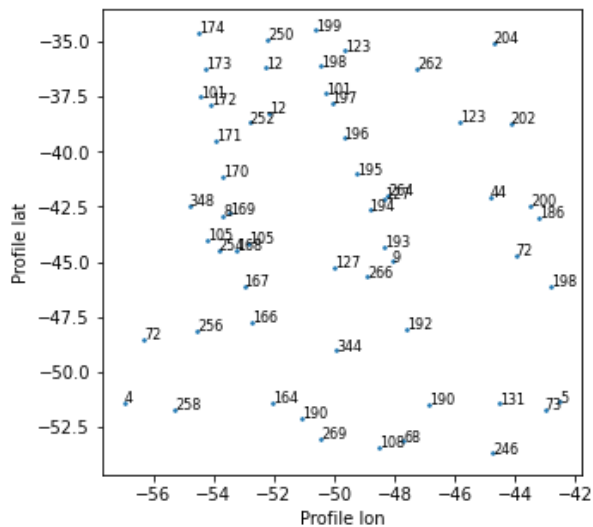

Fig. 3. Spatial distribution of profiles within Argyre Planitia during MY 34 $L_s$ 203° - 205° (dust storm)

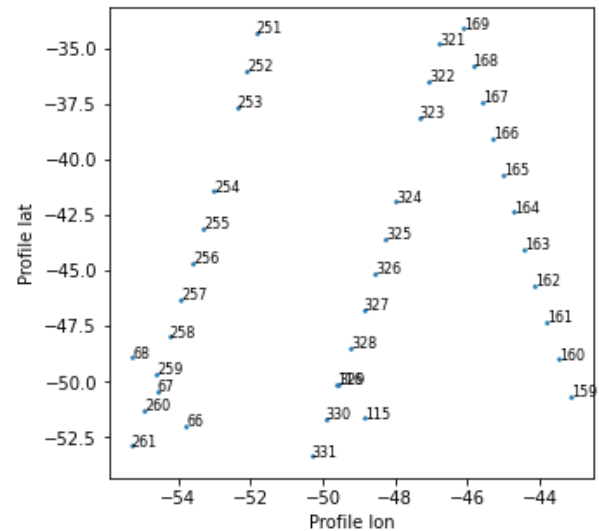

Fig. 5. Spatial distribution of profiles within Argyre Planitia during MY 35 $L_s$ 87° - 89° (clear conditions)

Table 4. MRO MCS data averaged over different altitude ranges during MY 35 $L_s$ 87° - 89° (clear conditions)

| Profile | $h$ [km] | $\overline{T}$ [K] | $\overline{P}$ [Pa] |
|---|---|---|---|
| All profiles | Full range | 149.692 | 54.946 |
|  | 0 - 40 | 167.980 | 143.12 |
|  | **0 - 20** | **171.943** | **267.03** |
|  | 0 - 15 | 171.309 | 324.88 |
| Sample profile * | Full range | 145.818 | 43.011 |
|  | 0 - 40 | 170.465 | 111.06 |
|  | 0 - 20 | 179.414 | 209.42 |
|  | 0 - 15 | 179.850 | 261.74 |

* **Id 324**: lat, lon: -41.909, -47.980; $L_s$: 87.39556°

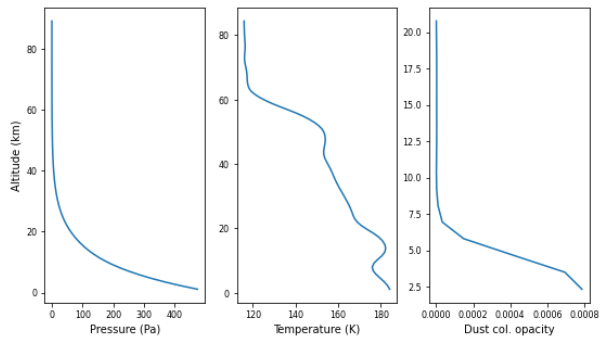

Fig. 6. Pressure, temperature and dust column opacity per altitude of the pressure surface above the surface point for sample profile Id 324 (during clear conditions)

*4.2. Atmospheric absorption coefficient estimation*

For this study, the Martian atmospheric absorption coefficient has been estimated using the high-resolution transmission molecular absorption database (HITRAN) API [37]. HITRAN contains a compilation of spectroscopic parameters used to predict the transmission and emission of light in the atmosphere. The results given by the HITRAN database for the Martian conditions have been evaluated in [38] using results obtained from the ExoMars Trace Gas Orbiter. While many challenges to address to improve the database were found, the results showed a reasonable performance for the current study.

Calculation of the absorption coefficient requires specification of the Martian atmospheric composition as well as the temperature and pressure conditions. As detailed in section 2.1, the atmospheric composition used in this study is a combination of $CO_2$, $N_2$, Ar, $O_2$, and CO in different percentages.

Pressure and temperature parameters are obtained from MRO MCS data compiled in section 4.1. In particular, the mean average temperature and pressure between 0 - 20 km altitude for all profiles during the MY 34 dust storm (Scenario 1: T = 197.399 K, P = 269.26 Pa) and MY 35 clear conditions (Scenario 2: T = 171.943 K, P = 267.03 Pa) respectively are considered for the analysis.

The frequency range analysed for this study is 100 - 700 THz, which encompases the visible, near-infrared and part of the mid-infrared regions of the electromagnetic spectrum [39]. Fig. 7 and Fig. 8 respectively compare the absorption coefficients within this frequency range for $H_2O$ and $CO_2$ molecules considering the atmospheric conditions in both Earth and Mars. Absorption coefficient values for CO and $O_2$ molecules in the atmosphere were significantly lower and would have a smaller impact on the total attenuation.

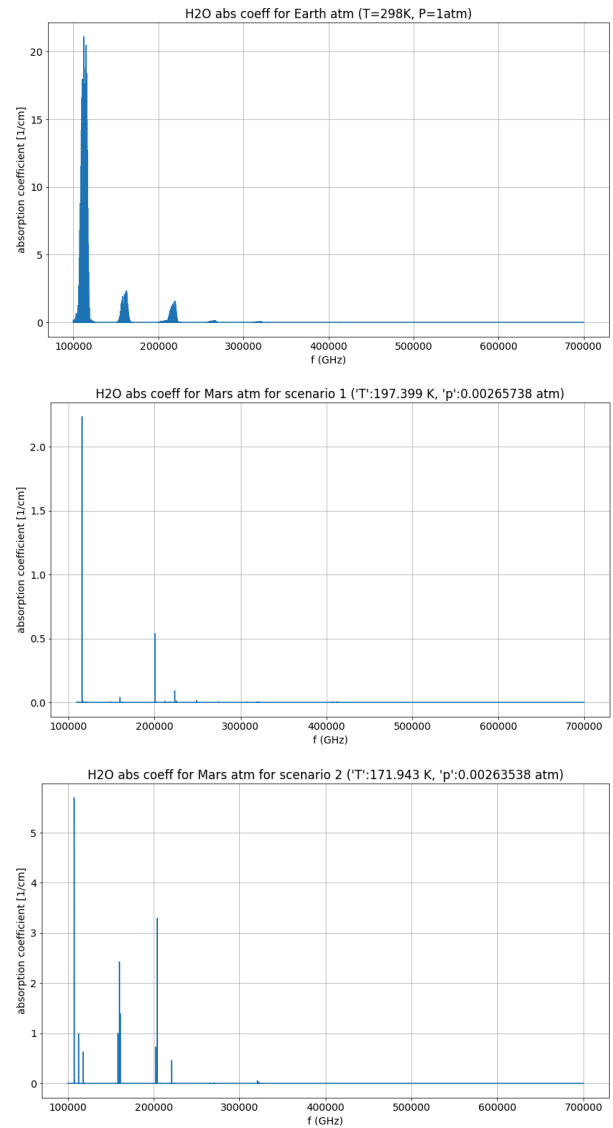

Fig. 7. $H_2O$ absorption coefficients calculated using HITRAN for nominal conditions on Earth and dust storm and clear atmosphere on Mars.

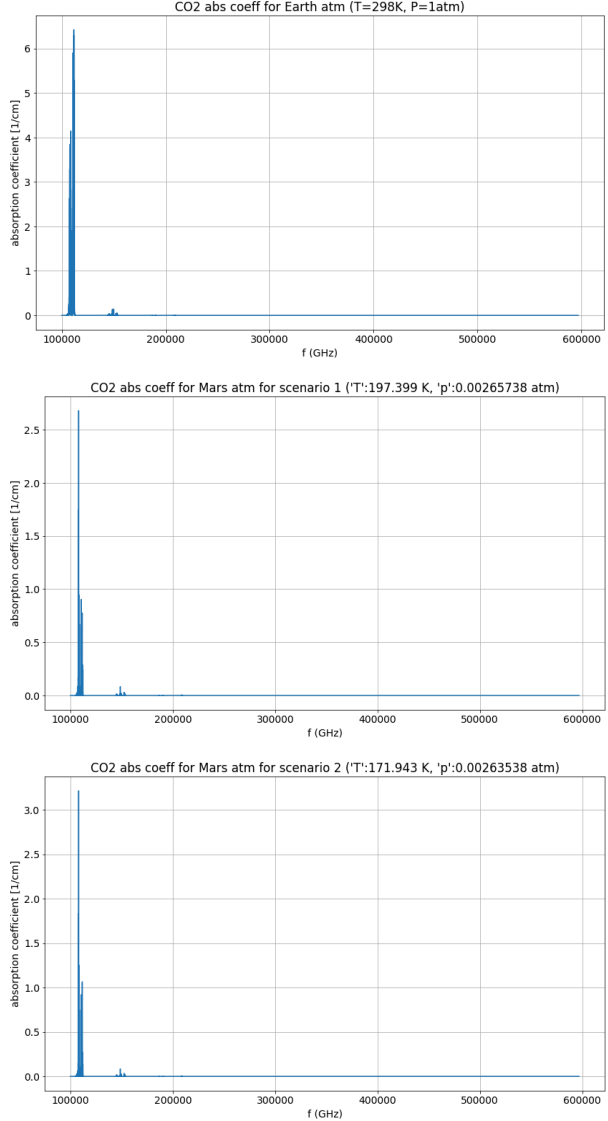

Fig. 8. $CO_2$ absorption coefficients calculated using HITRAN for nominal conditions on Earth and dust storm and clear atmosphere on Mars.

*4.3. Dust scattering coefficient estimation*

Scattering depends on factors such as the wavelength of the radiation, the abundance of particles and the distance travelled through the atmosphere by the emitted light. It occurs when particles in the atmosphere interact with and cause the electromagnetic radiation to be redirected from its original path [40]. There are three types of scattering that are distinguished by the relation between the particle size and the wavelength of the radiation: Rayleigh scattering occurs when the particles are very small compared to the wavelength, Mie scattering occurs when the particle size is comparable to the wavelength and non-selective or geometric scattering occurs when particles are much larger than the wavelength of the radiation.

The dominant feature of Martian atmospheric conditions that affects scattering (and thus should be primarily considered in calculation of the scattering coefficient) are suspended dust particles. The effect of dust particles on the attenuation of light is heavily dependent on the size of those particles and the wavelength of the transmission source. For this study, the considered effective radius range follows the references in [8] and is set to values between 1.4 and 2.5 micrometres.

The results shown in [8] and obtained in [41] regarding the extinction efficiency factor $Q_{ext}$ for martian dust reveal that it has an almost constant value of approximately 2.5 for the range of wavelengths under study. In addition, the same sources include the values for the single scattering albedo $w$ which is close to 1 for the larger part of the range of wavelengths and slowly decreases to 0.7 for shorter wavelengths.

The results shared in [42] provide more insight into the single scattering albedo $w$ for short wavelengths. The study shows that $w$ has an almost constant value of 0.97 for wavelengths from 700 to 2700 nm (corresponding to a frequency range of 111 - 428 THz) considering dust particles of 1.8 μm effective radius. $w$ follows an approximately linear decline from 700 nm to shorter wavelengths, reaching values around 0.76 for wavelengths close to 450 nm.

As the albedo is reasonably close to 1 in most of the frequency range considered, this study will follow a simplified approach where the absorption due to dust particles is assumed to be negligible (although its importance increases with increasing frequency). Thus, the extinction efficiency factor, $Q_{ext}$, is approximated to be equal to the scattering efficiency factor, $Q_{sca}$. To consider the effect of dust in the light transmission model it is necessary to calculate the scattering coefficient.

$$Q_{ext} = Q_{sca} + Q_{abs} \qquad (3)$$

Following the simplification for dust particles: $Q_{ext} \simeq Q_{sca}$.

The efficiency factor is related to the cross section ($\sigma_{sca}$) through the following expression from [43]:

$$\sigma_{sca} = Q_{sca} \pi R^2 \qquad (4)$$

The scattering coefficient definition is the cross-sectional area per unit volume of medium [44]:

$$sca.\ coeff. = \rho \sigma_{sca} \qquad (5)$$

As a reference, the order of magnitude of the dust concentration values considered in this paper are aligned with the estimations in [23], using 1.2 cm$^{-3}$ for dust density in the clear atmosphere and 17 and 12 cm$^{-3}$ for concentration during dust storms in MY25 and MY28 respectively. The results of these calculations are included in Table 5.

Table 5. Dust parameters in each scenario

| Scenario | $\rho$ [cm$^{-3}$] | Particle radius [m] | $\sigma$* [cm$^2$] | sca. coeff. [cm$^{-1}$] |
|---|---|---|---|---|
| Clear atmo. | 1.2 | 1.2e-6 | 1.131e-7 | 1.357e-7 |
| | | 1.8e-6 | 2.545e-7 | 3.054e-7 |
| | | 2.4e-6 | 4.524e7 | 5.429e-7 |
| MY28 dust storm | 12 | 1.2e-6 | 1.131e-7 | 1.357e-6 |
| | | 1.8e-6 | 2.545e-7 | 3.054e-6 |
| | | 2.4e-6 | 4.524e-7 | 5.429e-6 |
| MY25 dust storm | 17 | 1.2e-6 | 1.131e-7 | 1.922e-6 |
| | | 1.8e-6 | 2.545e-7 | 4.326e-6 |
| | | 2.4e-6 | 4.524e-7 | 7.691e-6 |

$Q_{sca}$ = 2.5 [8]
* $\sigma$ cross-section

The resultant values of the scattering coefficient are of a similar magnitude to existing literature. In [45], the authors estimated a scattering coefficient of about 2e-5 cm$^{-1}$ during the period of greater activity of the dust storm under study in the paper, where the mean radius of the particles was about 10 μm.

*4.4. Laser characteristics*

The selection of the laser in this study is driven by the frequency of the absorption bands obtained through analysis using HITRAN and considering the available literature regarding lasers in the THz band of interest. In general, the wavelength of a viable laser for optical communications depends on factors such as effects on the transmission due to the propagation through the channel and the availability and reliability of components for the transmitter and detector.

Different considerations on the optical signal transmission favour longer or shorter wavelengths beyond the benefits that higher frequencies provide in terms of throughput and bandwidth. The results in [46] show that in free conditions, energy density decreases by three orders of magnitude as the wavelength increases, which provides a motivation to select shorter wavelengths. However, the authors in [46] also explain that longer wavelengths are preferred due to less severe turbulence, scintillation, beam wander, and isoplanatic angle effects.

Analysing the values representing the absorption of $CO_2$, $CO$, $H_2O$ and $O_2$, it is possible to confirm a higher attenuation for specific wavelengths. Considering the dust storm and clear atmosphere scenarios analysed in this study, some spectral bands must be avoided to achieve reasonable link budgets. Ideally, the frequency band between 100 and 250 THz (~1.2 to 3 μm) should not be considered due to the peaks of absorption shown for $CO_2$, $H_2O$, and $CO$. In addition, the band between 400 and 500 THz (~0.6 to 0.75 μm) would be problematic due to absorption of CO, however, the values are much lower than the absorption of the other molecules. Therefore, the ranges between 250 and 400 THz (~0.75 to 1.2 μm) and 500 to 700 THz (~0.4 to 0.6 μm) would be the preferred bands for transmission. As an example, this study could consider a laser communication transmitter with 1 μm wavelength (i.e. approximately centred at 300 THz).

**5. Conclusions and future work**

This paper summarises the initial investigations performed as part of a project to develop a model for optical communications on Mars, exploring the effect of the Martian atmosphere and dust concentration in two different scenarios (during the MY34 dust storm and clear conditions in MY35) on an optical signal. This was done by deriving the absorption coefficient of the atmospheric gases for a range of frequencies and the scattering coefficient of Martian dust for a range of particulate sizes.

These calculations were informed both through literature review into Martian atmospheric conditions and optical communications, and analysis of MRO MCS data gathered in the Argyre Planitia.

The feasibility of a UAV communications architecture during Martian dust storms was briefly evaluated considering the stratospheric balloon design simulated in [19] as well as existing Earth-based balloon designs, leading to a concept for a two-step communications link assisted by a balloon UAV for data relay.

Following on from this work, the model will be improved by considering:

1. Future results from NASA's LCRD when available (and other relevant missions);
2. Inclusion of attenuation from $H_2O$ and $CO_2$ ice clouds;
3. The effect of changes in atmospheric temperature and pressure over altitude, using MCS data;
4. Estimation of dust concentration from satellite data for the specific event under study;
5. Consideration of additional sites and environmental conditions, such as day/night differences, to allow the model to be applied in a wider range of scenarios;
6. Calculation of received power for an optical signal in the Martian atmosphere;
7. Implementation of the model as a GNU Radio block, to allow for its use by the optical communications community.